\documentclass[fleqn,twoside]{article}
\usepackage{espcrc2}

\usepackage{graphicx}
\usepackage[figuresright]{rotating}

\newcommand{\AmS}{{\protect\the\textfont2
  A\kern-.1667em\lower.5ex\hbox{M}\kern-.125emS}}

\newcommand{\N}{\nonumber}

\title{Threshold expansion of massive coloured 
particle cross sections}

\author{
 M. Beneke\address[AACHEN]{
                 Institut f\"ur Theoretische Teilchenphysik und Kosmologie,
                 RWTH Aachen University,\\ 
                 \hspace*{0.16cm} D--52056 Aachen, Germany}
                 \thanks{Speaker. To appear in: Proceedings of the 
                 10th DESY Workshop on Elementary Particle Theory: 
                 Loops and Legs in Quantum Field Theory 2010,
                 W\"orlitz, Germany, 25-30 Apr 2010.
                 Preprint Nos. TTK-10-43, SFB/CPP-10-64, IPPP-10-58, 
                 DCPT-10-116, FR-PHENO-2010-025.},
 P. Falgari\address{IPPP, Department of Physics, Durham, Durham DH1 3LE, 
                    England},
 S. Klein\addressmark[AACHEN],
 C. Schwinn\address{Albert--Ludwigs Universit\"at Freiburg, 
                    Physikalisches Institut, 
                    D--79104 Freiburg, Germany},
  }

\begin{document}

\begin{abstract}
\noindent
Pair production of massive coloured particles in hadron collisions is 
accompanied by potentially large radiative corrections related to 
the suppression of soft gluon emission and enhanced Coulomb exchange 
near the production threshold. We recently developed a framework to 
sum both series of corrections for the partonic cross section 
using soft-collinear and non-relativistic effective theory. If it 
can be argued that the resummed cross section approximates the 
complete result over a significant kinematic range, an improvement 
of the hadronic cross section results, even when the production is 
not kinematically constrained to the threshold. 
This is discussed here for the case of top quark production.
\vspace{1pc}
\end{abstract}

\maketitle
%
%
%
\section{Introduction}

\noindent
The rediscovery of the top quark at the Large Hadron Collider (LHC) 
in the near future will mark the beginning of an era of precision 
studies of the properties of the heaviest of all quarks. There is 
therefore currently much interest in predicting the production 
cross section and invariant mass distribution 
precisely~\cite{Moch:2008qy,Cacciari:2008zb,Kidonakis:2008mu,Langenfeld:2009wd,Hagiwara:2008df,Kiyo:2008bv,Ahrens:2010zv}, beyond 
the fixed-order NLO result~\cite{Nason:1987xz}, by
extending and updating earlier calculations including soft gluon 
resummation~\cite{Catani:1996dj,Berger:1996ad,Kidonakis:1996zd,Bonciani:1998vc},
 using recent results on NNLL resummation for massive particles~\cite{Kidonakis:2009ev,Mitov:2009sv,Becher:2009kw,Beneke:2009rj,Czakon:2009zw,Ferroglia:2009ep}.

In pair production of coloured particles an additional power-like threshold 
enhancement, formally stronger than the logarithmic enhancement 
related to soft gluons, arises due to the colour-Coulomb force. The 
question whether the standard resummation formalism must be modified 
in the presence of the strong Coulomb interaction, and whether 
both types of enhancements can be simultaneously resummed to all 
orders was not addressed in the above papers, which usually included 
the Coulomb correction at the one-loop level or assumed 
factorization of soft gluon and Coulomb effects. In 
\cite{Beneke:2009rj,Beneke:2010gi} we extended the momentum-space 
formalism for resummation \cite{Becher:2007ty} based on soft-collinear 
effective theory (SCET) to the case of pair production, showing factorization 
of the partonic cross section of the form 
\begin{eqnarray}
\hat\sigma(\beta,\mu)
&=& \sum_a \sum_{i,i'}H_{ii'}^a(m_t,\mu)
\nonumber\\
&&\hspace*{-1.5cm}\times\, \int d \omega\;
\sum_{R_\alpha}\,J_{R_\alpha}^a(E-\frac{\omega}{2})\,
W^{a,R_\alpha}_{ii'}(\omega,\mu),
\label{factform}
\end{eqnarray}
with the top-quark velocity $\beta=(1-4 m_t^2/\hat s)^{1/2}$, 
$E=\sqrt{\hat{s}}-2m_t \approx m_t\beta^2$ and $\sqrt{\hat{s}}$ being 
the partonic cms energy.
Eq. (\ref{factform}) contains a multiplicative short-distance coefficient 
$H_{ii'}^a$ in each colour (in higher orders also spin) configuration 
labelled by the irreducible representation $R_\alpha$, 
and a convolution of soft functions $W^{a,R_\alpha}_{ii'}$ with 
functions $J_{R_\alpha}^a$, which contain Coulomb exchange to all 
orders. The factorization in this form implies that the soft and Coulomb
corrections can both be summed.

The oral presentation covered a detailed discussion of the 
above factorization formula, our results for squark-antisquark 
production at the next-to-leading logarithmic order, 
and for top-quark pair production. Since the former have 
already been documented in other proceedings articles
\cite{Beneke:2009nr,Beneke:2010gm}, we focus here on our 
preliminary results for top quarks.

%
%
%
\section{Top quark production}
\noindent 
First we consider the inclusive partonic cross section
for $t\overline{t}$-production, denoted by 
$\hat{\sigma}_{t\overline{t}}(\beta)$. The series of 
enhanced radiative corrections can be represented 
parametrically as 
\begin{eqnarray}
\hat{\sigma}_{t\overline{t}}(\beta)&=&
\hat{\sigma}_{t\overline{t}}^{(0)}\sum_{k=0}
\,\Bigl(\frac{\alpha_s}{\beta}\Bigr)^k
\exp \Bigl[\underbrace{\ln \beta \,g_0(\alpha_s\ln\beta)}_{({\rm LL})}
\nonumber \\ 
&& \hspace{-10mm}
+\underbrace{g_1(\alpha_s\ln\beta)}_{({\rm NLL})}
+\underbrace{\alpha_sg_2(\alpha_s\ln\beta)}_{({\rm NNLL})}
+\ldots\Bigr]
\nonumber \\ 
&& \hspace{-10mm}
\times \,\Bigl\{1{\rm (LL,NLL)};\alpha_s,\beta{\rm (NNLL)};\ldots\Bigr\}\,, 
\label{sigttbar}
\end{eqnarray}
where $\hat{\sigma}_{t\overline{t}}^{(0)}$ is the Born cross section. 
In terms of the fixed-order expansion the different orders 
of resummation refer to 
\begin{eqnarray}
 {\rm LL:} \!\!\!\!&& \!\!\!\!
              \alpha_s\Bigl\{\frac{1}{\beta},\ln^2\beta\Bigr\};~
              \alpha^2_s\Bigl\{\frac{1}{\beta^2},
                        \frac{\ln^2\beta}{\beta},\ln^4\beta\Bigr\} ;\ldots
\nonumber \\
 {\rm NLL:}\!\!\!\!&&\!\!\!\!
              \alpha_s \ln\beta;~
              \alpha^2_s\Bigl\{\frac{\ln\beta}{\beta},
                        \ln^3\beta\Bigr\};\ldots
\end{eqnarray}
etc. In obtaining the hadronic total cross section the 
partonic cross section is integrated over all $\beta$ up to 
the kinematic constraint $\beta_{\rm max}=(1-4 m_t^2/s)^{1/2}$, weighted 
by the parton luminosity. The threshold expansion is strictly valid for the 
hadronic cross section only for
high masses $2 m_t \rightarrow s$ such that $\beta_{\rm max}\to 0$,
but certainly not for tops at the Tevatron and the LHC with 
$\sqrt{s}=7\,$TeV and higher energy. Nevertheless, one sometimes 
finds that the threshold expansion provides a reasonable approximation 
even outside its domain of validity, so that it can be useful to 
include the threshold limit of higher-order terms in the perturbative 
expansion. 

\begin{figure}[htb]
\begin{center}
\includegraphics[angle=0, width=7.0cm]{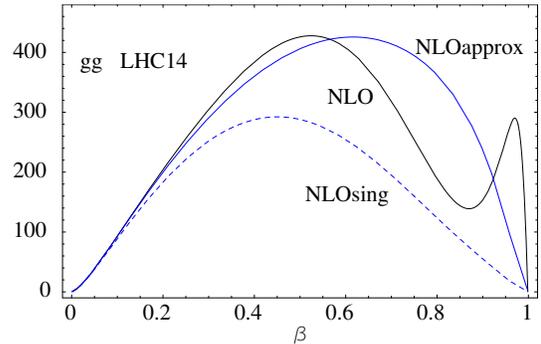}
\end{center}
\vspace*{-13mm}
\caption{\sf $\beta$--integrand  (in pb) for different approximations 
to the NLO correction to the hadronic cross section in the $gg$-channel 
for $t\bar t$ production in $pp$ collisions at $\sqrt{s}=14\,$TeV.} 
\label{NLOapprox}
\end{figure}

It is therefore interesting to investigate this issue 
at the next-to-leading order (NLO), where the exact result is known. 
Here, the $t\bar t$ invariant mass distribution 
peaks at about 380 GeV, corresponding to $\beta \approx 0.4$, but the average 
$\beta$ is even larger, see Table~\ref{table:1} for the 
gluon-gluon production channel. To check whether the threshold 
expansion can be a reasonable approximation we show in 
Figure \ref{NLOapprox} the $\beta$-integrand of the NLO correction 
to the hadronic cross section in the $gg$-channel 
\begin{equation}
\frac{d\sigma_{t\bar t}}{d\beta} = 
  \frac{8\beta m_t^2}{s(1-\beta^2)^2}\,{\cal L}_{gg}(\beta)\,
  \hat{\sigma}_{t\bar t}^{(gg)}(\beta)\,, 
\end{equation}
where ${\cal L}_{gg}$
is the gluon parton luminosity. The Figure displays the full NLO 
result \cite{Nason:1987xz,Czakon:2008ii} and compares to it 
the approximation NLO$_{\rm sing}$ where only the singular terms $1/\beta$, 
$\ln^2\beta$, $\ln \beta$ in the threshold expansion of the 
NLO correction (normalized to the Born cross section) are  
kept and to NLO$_{\rm approx}$, which includes in addition the 
constant term $\beta^0$ in the $\beta$ expansion, see 
for instance Eq.~(A.2) of Ref.~\cite{Beneke:2009ye}. 
(We use $m_t=173.1$ GeV and set the 
renormalization and factorization scale equal to $\mu_r=\mu_f=m_t$.) 
We see that NLO$_{\rm approx}$ provides a good approximation 
up to $\beta\approx 0.6$. In Table~\ref{table:1} we show 
the corresponding results for the integrated NLO cross sections with the 
approximations to the NLO correction as discussed above, now 
including all partonic production channels.

We therefore make the assumption that 
the threshold expansion provides a good approximation 
for the integral over all $\beta$, i.e. the total hadronic 
cross section. As shown, this works reasonably well 
for the $gg$-channel at NLO (but less well for the 
$q\overline{q}$-channel). We expect that this approximation becomes 
better at NNLO, because the average is dominated by smaller 
$\beta$ as the order increases due to the existence of more 
singular terms. 

\begin{table}[tb]
\caption{NLO results for different approximations}
\label{table:1}
\newcommand{\m}{\hphantom{$-$}}
\newcommand{\cc}[1]{\multicolumn{1}{c}{#1}}
\renewcommand{\tabcolsep}{0.9pc} 
\renewcommand{\arraystretch}{1.2} 
\begin{tabular}{@{}lllll}
\hline
                            & Tev.     & LHC7  & LHC14 \\
\hline
$\langle\beta\rangle_{gg,\,\rm NLO}$     & 0.41     & 0.49  & 0.53  \\
LO                          & 5.25     & 101.9 & 562.9 \\
NLO                         & 6.50     & 149.9 & 842.2 \\
NLO$_{\rm sing}$           & 6.76     & 138.8 & 751.2 \\
NLO$_{\rm approx}$         & 7.45     & 159.0 & 867.6 \\
\hline
\end{tabular}\\[2pt]
Top pair production cross section in pb at the Tevatron (Tev.) and LHC with 
$\sqrt{s}=7\,$TeV and $14\,$TeV; MSTW2008nnlo PDFs \cite{Martin:2009iq}.
\end{table}

Let us now turn to our results for NLL resummation of the total 
hadronic cross section for $t\overline{t}$-production. As mentioned above, 
we apply the SCET formalism \cite{Becher:2007ty}, rather than working 
in Mellin-space \cite{Bonciani:1998vc}. Motivated by our previous 
findings, we integrate the resummed cross section over 
all values of $\beta$ and do not switch off the threshold 
resummation outside its formal domain of validity.
We include soft gluon and Coulomb gluon resummation and match 
the NLL resummed cross section to the full NLO result. Thus our 
first approximation, NLL+NLO, is given by
\begin{eqnarray}
 \sigma^{{\rm NLL+NLO}}_{t\overline{t}}
   &=& \sigma^{{\rm NLL}}_{t\overline{t}}
      -\sigma^{{\rm NLL}}_{t\overline{t}}\Big|_{\rm NLO}
      +\sigma^{{\rm NLO}}_{t\overline{t}}\,,
\end{eqnarray}
which is NLO exact but further includes all NLL terms beyond NLO. 
Here and in the following, 
$\sigma^{{\rm NLL}}_{t\overline{t}}\Big|_{{\rm N(N)LO}}$ is given by
$\sigma^{{\rm NLL}}_{t\overline{t}}$ expanded in $\alpha_s$
up to N(N)LO accuracy. 
Next we define NNLO$_{\rm approx}$ to be the sum of the 
the exact NLO result plus all singular terms in $\beta$ at $O(\alpha_s^2)$, 
which were determined in Ref.~\cite{Beneke:2009ye}. 
Finally, our third and best approximation is given by
\begin{eqnarray}
 \sigma^{{\rm NLL+NNLO_{\rm approx}}}_{t\overline{t}}
   &=& \sigma^{{\rm NLL}}_{t\overline{t}}
      -\sigma^{{\rm NLL}}_{t\overline{t}}\Big|_{{\rm NNLO}}
\N\\ &&
      +\sigma^{{\rm NNLO_{\rm approx}}}_{t\overline{t}}~.
\end{eqnarray}

\begin{figure}[tb]
\begin{center}
\includegraphics[angle=0, width=7.5cm]{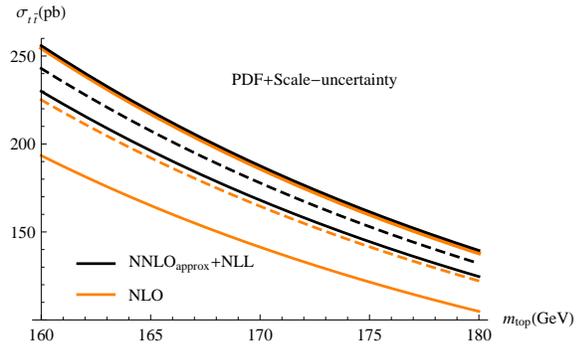}
\end{center}
\vspace*{-10mm}
\caption{\sf $\sigma_{t\bar t}$ in $pp$ collisions at 
$\sqrt{s}=7\,$TeV shown in dependence on $m_t$. 
The dashed lines are the central values, while the solid
lines represent the combined MSTW2008nnlo PDF and scale uncertainty.}
\label{ScalePdfUn}
\vspace*{-3mm}
\end{figure}

\begin{table*}[htb]
\caption{Preliminary results for NLL resummation}
\label{table:2}
\newcommand{\m}{\hphantom{$-$}}
\newcommand{\cc}[1]{\multicolumn{1}{c}{#1}}
\renewcommand{\tabcolsep}{0.8pc} 
\renewcommand{\arraystretch}{1.0} 
\begin{tabular}{@{}llllll}
\hline
 &       & Tevatron & LHC7 & LHC10 & LHC14 \\
\hline \\
NLO & MSTW08
 & $\phantom{0}6.50^{\,+0.32+0.33}_{\,-0.70-0.24}$
 & $\phantom{0}150^{\,+18+8}_{\,-19-8}$
 & $\phantom{0}380^{\,+44+17}_{\,-46-17}$
 & $\phantom{0}842^{\,+97+30}_{\,-97-32}$ \vspace{1mm} \\ 
 & ABKM09
 & $\phantom{0}6.43^{\,+0.23+0.15}_{\,-0.61-0.15}$
 & $\phantom{0}122^{\,+13+7}_{\,-15-7}$
 & $\phantom{0}322^{\,+36+15}_{\,-38-15}$
 & $\phantom{0}738^{\,+81+27}_{\,-83-27}$  \vspace{2mm} \\ 
NLL+NLO & MSTW08
 & $\phantom{0}6.54^{\,+0.98+0.33}_{\,-0.38-0.24}$
 & $\phantom{0}151^{\,+24+8}_{\,-14-8}$
 & $\phantom{0}381^{\,+60+17}_{\,-36-17}$
 & $\phantom{0}845^{\,+131+30}_{\,-81-32}$ \vspace{1mm} \\ 
 & ABKM09
 & $\phantom{0}6.46^{\,+0.89+0.15}_{\,-0.35-0.15}$
 & $\phantom{0}122^{\,+19+7}_{\,-11-7}$
 & $\phantom{0}323^{\,+49+15}_{\,-30-15}$
 & $\phantom{0}741^{\,+110+27}_{\,-69-27}$ \vspace{2mm} \\ 
NNLO$_{\rm approx}$ & MSTW08
 & $\phantom{0}7.13^{\,+0.00+0.36}_{\,-0.33-0.26}$
 & $\phantom{0}162^{\,+3+9}_{\,-3-9}$
 & $\phantom{0}407^{\,+11+17}_{\,-5-18}$
 & $\phantom{0}895^{\,+29+31}_{\,-7-33}$ \vspace{1mm} \\ 
 & ABKM09
 & $\phantom{0}7.01^{\,+0.06+0.18}_{\,-0.36-0.18}$
 & $\phantom{0}132^{\,+2+8}_{\,-2-8}$
 & $\phantom{0}345^{\,+8+16}_{\,-3-16}$
 & $\phantom{0}785^{\,+22+29}_{\,-6-29}$ \vspace{2mm} \\ 
NNLO$_{\rm approx}$ + NLL & MSTW08
 & $\phantom{0}7.13^{\,+0.08+0.36}_{\,-0.41-0.26}$
 & $\phantom{0}162^{\,+2+9}_{\,-1-9}$
 & $\phantom{0}407^{\,+9+17}_{\,-2-18}$
 & $\phantom{0}895^{\,+23+31}_{\,-4-33}$ \vspace{1mm} \\ 
 & ABKM09
 & $\phantom{0}7.00^{\,+0.13+0.18}_{\,-0.44-0.18}$
 & $\phantom{0}132^{\,+1+8}_{\,-1-8}$
 & $\phantom{0}345^{\,+6+16}_{\,-1-16}$
 & $\phantom{0}784^{\,+17+29}_{\,-3-29}$ \vspace{2mm} \\ 
\hline
\end{tabular}\\[2pt]
Top-quark pair production cross section in pb at the Tevatron and in 
$pp$ collisions at $\sqrt{s}=7, 10, 14\,$TeV. 
MSTW2008nnlo and ABKM09 PDFs. Error includes
scale variation $\mu_i/2,\ldots,2\mu_i$ for $i=f,s,h$ (first number)
and PDF error (second error).
\end{table*}

Within the SCET formalism, there are several scales involved 
in the resummed cross section. The hard scale $\mu_h$, which is the 
scale of the hard matching coefficients,
is taken to be equal to $2m_t$.
The soft scale $\mu_s$ is determined by minimizing the one-loop soft 
corrections. The Coulomb scale is chosen as
$\mu_C={\rm Max}[\alpha_s(\mu_C)C_Fm_t,2m_t\beta]$. In order 
to estimate the scale uncertainty, we vary the scales 
$\mu_f, \mu_s$ and $\mu_h$ by a factor of 2.
Our results for the $t\bar{t}$ cross section are summarized in 
Table~\ref{table:2} and Figure~\ref{ScalePdfUn}, using the MSTW2008nnlo 
parton distribution functions (PDFs)~\cite{Martin:2009iq}, 
and the ABKM09nnlo PDFs \cite{Alekhin:2009ni}. The two uncertainties 
shown explicitly in the Table
stem from the variation of the scales (first error) and from the 
PDF uncertainty (second error). The two PDF sets lead to cross sections 
consistent within their stated uncertainties for the Tevatron,
but for LHC energies the results for the MSTW08 set are larger by
$2\sigma-3\sigma$, depending on the approximation one considers. 
This is due to a larger value of $\alpha_s(M_Z)$  and a 
larger value of the gluon PDF in the
partonic threshold region $\hat{s} \simeq  4 m_t^2$ for the MSTW08 PDF set.
We can directly compare our results for the  ${\rm NLL+NNLO_{\rm approx}}$ 
approximation with Ref.~\cite{Ahrens:2010zv}. The corresponding result  
is called $\sigma_{{\rm NNLO},\beta{\rm -exp.+potential}}$ there
and is in full agreement for $\mu_f=m_t$.
We note that adding NLL resummation to ${\rm NNLO_{\rm approx}}$ 
has only a permille effect. The scale dependence is reduced 
drastically once the singular terms at $O(\alpha_s^2)$ are included, 
from about $10\%$ for the NLO result to $(1-2)\%$ for 
NNLO$_{\rm approx}$+NLL. In Figure \ref{ScalePdfUn} we plot 
the dependence of the resummed cross section on the top mass 
and compare to the NLO result for $\sqrt{s}=7\,$TeV. The resummed 
cross section is enhanced by about $10\%$, and the combined
PDF and scale uncertainty is reduced by roughly $50\%$.

\section{Summary}

\noindent
We presented a progress report of our work on the combined soft and 
Coulomb resummation for top-quark pair production in hadron 
collisions. Including the singular terms near threshold at $O(\alpha_s^2)$ 
leads to an enhancement of the cross section and a significant reduction 
of the scale dependence. Summing NLL logarithms and the leading 
Coulomb corrections beyond this order is a minor effect. The 
complete NNLL resummation can be readily performed in the 
SCET plus NRQCD formalism and is in progress.

\section*{Acknowledgments}
\noindent 
S.K. thanks S. Alekhin for use\-ful discussions. 
This work has been supported in part 
by the DFG Sonder\-for\-schungs\-bereich/Trans\-regio~9 
``Computergest\"utzte Theoreti\-sche Teilchenphysik''.


\end{document}